\documentclass{article}

\usepackage{PRIMEarxiv}

\usepackage[utf8]{inputenc} 
\usepackage[T1]{fontenc}    
\usepackage{hyperref}       
\usepackage{url}            
\usepackage{booktabs}       
\usepackage{amsfonts}       
\usepackage{nicefrac}       
\usepackage{microtype}      
\usepackage{lipsum}
\usepackage{graphicx}
\usepackage{colortbl}
\usepackage[table,xcdraw]{xcolor}
\usepackage{multicol,multirow}
\graphicspath{{media/}}     

\definecolor{RowHeader}{HTML}{A9C4EB}
\definecolor{RowAlt}{HTML}{DDF5FF}
  
\title{Managing Diabetic Retinopathy with Deep Learning: A Data Centric Overview
}

\author{
  Shramana Dey \\
  Indian Statistical Institute \\
  Kolkata, West Bengal, India \\
  \And
  Zahir Khan \\
  Indian Statistical Institute \\
  Kolkata, West Bengal, India \\
  \And
  T. A. PramodKumar \\
  Dr.\ Mohan’s Diabetes Specialties Centre \\
  Madras Diabetes Research Foundation\\
  Chennai, Tamil Nadu, India \\
  \And
  B.\ Uma Shankar \\
  Indian Statistical Institute \\
  Kolkata, West Bengal, India \\
  \And
  Ashis K.\ Dhara\\
  Department of Electrical Engineering\\
  National Institute of Technology\\
  Durgapur, West Bengal, India\\
  \And
  Ramachandran Rajalakshmi\\
  Dr.\ Mohan’s Diabetes Specialties Centre \\
  Madras Diabetes Research Foundation\\
  Chennai, Tamil Nadu, India \\
  \And
  Rajiv Raman\\
  Department of Vitreo-Retina\\
  Vision Research Foundation, Sankara Nethralaya\\
  Chennai, Tamil Nadu, India \\
  \And
  Sushmita Mitra \\
  Indian Statistical Institute \\
  Kolkata, West Bengal, India \\
}

\begin{document}

\maketitle

\begin{abstract}
Diabetic Retinopathy (DR) is a serious microvascular complication of diabetes, and one of the leading causes of vision loss worldwide. Although automated detection and grading, with Deep Learning (DL), can reduce the burden on ophthalmologists, it is constrained by the limited availability of high-quality datasets. Existing repositories often remain geographically narrow, contain limited samples, and exhibit inconsistent annotations or variable image quality; thereby, restricting their clinical reliability. This paper presents a comprehensive review and comparative analysis of fundus image datasets used in the management of DR. The study evaluates their usability across key tasks, including binary classification, severity grading, lesion localization, and multi-disease screening. It also categorizes the  datasets by size, accessibility, and annotation type (such as image-level, lesion-level, and multi-disease). Finally, a recently published dataset is presented as a case study to illustrate broader challenges in dataset curation and usage. The review consolidates current knowledge while highlighting persistent gaps such as the lack of standardized lesion-level annotations and longitudinal data. It also outlines recommendations for future dataset development to support clinically reliable and explainable solutions in DR screening.
\end{abstract}


\section{Introduction}
\label{sec:intro}
Diabetes Mellitus (DM) is a chronic metabolic disorder that affected an estimated 537 million adults worldwide in 2021, a number that is expected to increase to 643 million by 2030 and 783 million by 2045 \footnote{Diabetes Atlas, 10th ed., International Diabetes Federation, Brussels, Belgium, 2021. https://diabetesatlas.org/resources/idf-diabetes-atlas-2025/}. One of the most serious microvascular complications of diabetes is Diabetic Retinopathy (DR), which affects approximately a third of the population with diabetes and remains the leading cause of irreversible visual impairment worldwide \cite{akhtar2025deep,raman2022prevalence}. 
According to the World Health Organization (WHO), the number of people affected by diabetes in India was 31.7 million in 2000.  India currently has more than 100 million people with diabetes; the highest in any country worldwide \cite{anjana2023metabolic,raman2009prevalence}.  DR is expected to affect nearly two\textcolor{blue}{-}thirds of individuals with type 2 diabetes and almost all with type 1 diabetes in the long run \cite{gadkari2016prevalence}.
Regions with a high prevalence of diabetes experience an elevated burden of DR worldwide. The All India Ophthalmological Society (AIOS) launched a nationwide DR screening program in 2014--the first non-governmental initiative of its kind in India--to assess its prevalence and risk factors \cite{gadkari2016prevalence}. Region-specific studies \cite{rema2005prevalence,nirmalan2004prevalence} reported considerable variation in DR prevalence. The Chennai Urban Rural Epidemiology Study (CURES) \cite{rema2005prevalence} documented 17.6\% in urban populations, while the Aravind Comprehensive Eye Study \cite{nirmalan2004prevalence} reported 10.5\% in rural South India.

\begin{figure}[http]
    \centering
    \includegraphics[width=0.8\columnwidth]{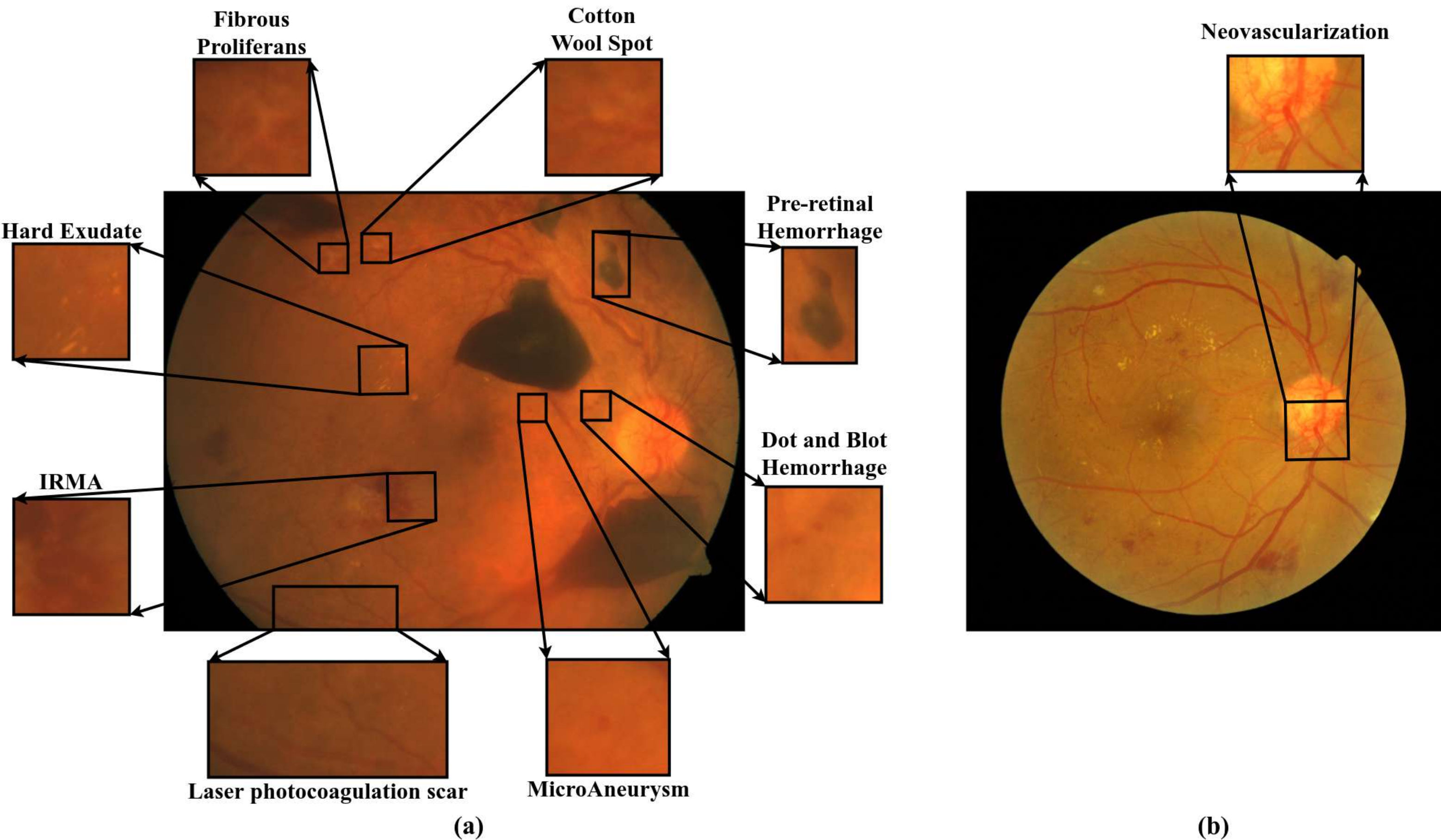}
    \caption{Sample fundus images depicting the eye pathologies in lesion formation for DR, like (a) microaneurysms (MA), hemorrhages (HE), hard exudates (HX), soft exudates (SX), intraretinal microvascular abnormalities (IRMA), among others, and (b) neovascularization (NV).}
    \label{fig:DRpathology}
\end{figure}

\begin{figure}[htbp]
\centering
\includegraphics[width=0.8\columnwidth]{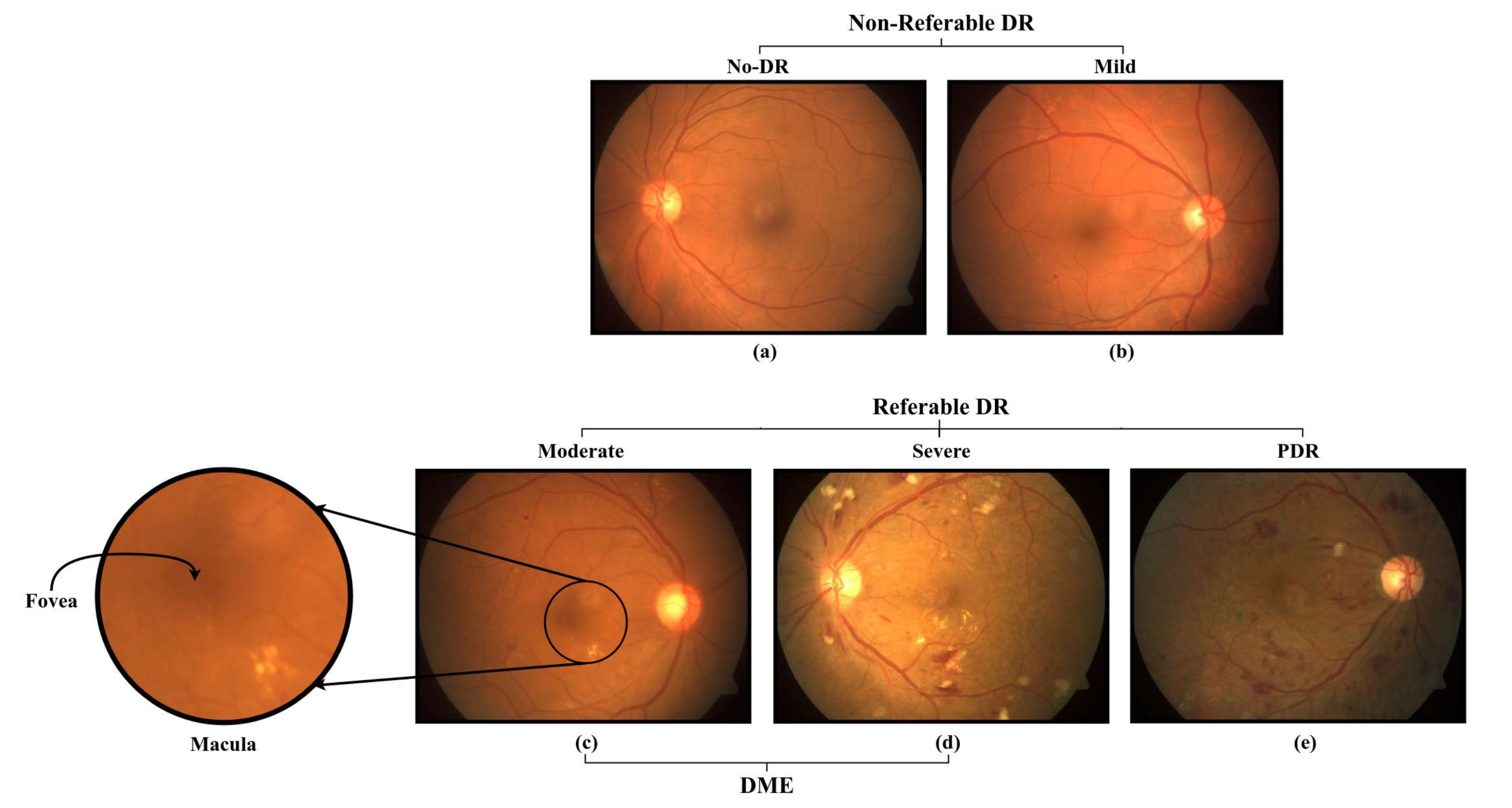}
\caption{Representative fundus images, illustrating the severity grades of DR. (a) Healthy case, and the (b)--(e) DR-affected cases.}
\label{fig:DRSeverity}
\end{figure}

The progression of DR is typically graded from mild Non-Proliferative DR (NPDR) to moderate and severe NPDR, culminating in Proliferative Diabetic Retinopathy (PDR). Diabetes-induced microvascular damage produces a spectrum of retinal lesions, as shown in Fig. \ref{fig:DRpathology}. The first characteristic includes microaneurysms (MA), which form as small protrusions of the capillary walls. Intraretinal hemorrhages (HE), such as dot and blot lesions, are produced in the deeper layers. Infrequently, flame-shaped hemorrhages (HE) form in the superficial nerve fiber layer. The leakage of lipids and proteins generates hard exudates (HX), while localized ischemia causes cotton wool spots (SX). Progressive capillary non-perfusion results in venous beading and intraretinal microvascular abnormalities (IRMA), which characterize severe NPDR. Neovascularization (NV) defines PDR in its advanced stage. Fluid accumulation in the macula produces diabetic macular edema (DME), which can occur at any stage of the disease and is one of the leading causes of vision loss \cite{wilkinson2003proposed}. Referable DR (RDR), typically defined as moderate NPDR or worse and/or the presence of DME warrants prompt ophthalmic referral. Fig. \ref{fig:DRSeverity} illustrates representative fundus images in this severity spectrum. DR and DME are the leading causes of blindness. Thus, accurate early detection of DME and DR, individually or jointly, is crucial for the diagnosis of irreversible but preventable blindness. 

Screening for DR involves the use of both invasive and non-invasive imaging. Invasive procedures, such as fluorescein angiography and Optical Coherence Tomography (OCT) angiography, provide detailed vascular information. However, they remain impractical on a large scale due to dye requirements, specialized equipment, and potential side effects. Non-invasive color fundus photography offers a safe, rapid and cost-effective alternative, making it the preferred modality for population-wide programs. Its widespread availability and compatibility with automated systems have placed fundus imaging datasets as the foundation for the development of effective DR screening tools for early detection.
Conventional DR screening relies on the acquisition of retinal fundus photographs, interpreted by trained graders or ophthalmologists. However, this process is labor intensive and time consuming for large-scale implementation, particularly when used in remote regions with limited availability of specialists. 

As classical image processing techniques are based on hand-crafted features, hard-threshold values, and rigid rules, handling the varied range of image quality poses a great challenge \cite{kulkarni2025convolutional,cutur2025multi}. 
Being less adaptive, they fail to generalize across diverse conditions and perform poorly when evaluated on a large dataset. Manual processing also entails the ill effects of human bias and slower speed.
Deep Learning (DL) methods have emerged as promising solutions for automated DR detection, classification and segmentation, showing performance on par with retinal specialists in several studies \cite{shen2017deep}\nocite{senapati2024artificial,dey2023multi,teng2024medidrnet}-\cite{dey2024adaptive}. 
The use of DL in Convolution Neural Network (CNN)- or transformer-based frameworks, combined with pre-trained weights, offers a faster and efficient solution to automated screening with a significant improvement in diagnostic accuracy. DL architectures that leverage backbone models such as VGG \cite{vgg}, ResNet \cite{resnet}, InceptionNet \cite{inception}, EfficientNet \cite{efficient}, and DenseNet \cite{dense}, have consistently shown promising results in a wide range of computer vision and medical imaging tasks. In recent years, Vision Transformer (ViT) \cite{vit} based architectures have gained significant attention in the field of medical image analysis, demonstrating competitive performance w.r.t. CNNs. 

 The reliability and clinical utility of these systems depend critically on the quality, size, and representativeness of the datasets used for training. Public datasets such as EyePACS \cite{diabetic-retinopathy-detection}, Messidor \cite{messidor}, DDR \cite{li2019diagnostic}, and IDRiD \cite{idrid} have been used with significant performance improvement. However, there are limitations in terms of demographic diversity, sample size, and annotation detail. High-quality datasets, with standardized annotations, play a central role in building generalizable models; thereby, enabling fair benchmarking, and supporting and evaluating performance for clinical translation. This review systematically builds on the available fundus image datasets, for the early detection of DR through screening based on automated DL strategies. It outlines their characteristics, strengths, and limitations, while providing a comprehensive perspective to guide the future development of automated screening tools.

The remainder of the manuscript is organized as follows. Section \ref{sec:dl_task} describes the basics of DL, and outlines some automated screening tasks with their clinical implications. Section \ref{sec:background} discusses the challenges of obtaining, curating, and annotating datasets for DR screening. It also reviews the major publicly available DR datasets, providing a detailed account of their characteristics, annotation types, and suitability for these tasks. Section \ref{sec:sanmod} highlights performance trends in the classification of DR, reported on a recently published dataset as a case study, to summarize key insights in the screening process using DL. Finally, Section \ref{sec:conclusion} concludes the article.

\section{Automated Screening of Diabetic Retinopathy}
\label{sec:dl_task}

Automated deep learning systems for the screening of DR from fundus images have advanced rapidly in recent years. In general, these tasks fall into three categories {\it ie,} classification, segmentation, and detection; each targeting different aspects of the screening process with distinct utilities. 

\subsection{Deep learning}

The Visual Geometry Group (VGG) \cite{vgg} models represent one of the first CNN architectures to exhibit promising results in computer vision tasks. The VGG16 consists of 16 layers, with a simple architecture based on stacked 3$\times$3 convolutional layers followed by fully connected layers. The simplicity of VGG networks allowed for efficient hierarchical feature extraction, making it effective for various image classification tasks. While the shallower layers captured simple local features (such as edges), the deeper layers handled complex semantic features. This made the VGG suitable for smaller datasets such as IDRiD \cite{idrid}.

However, deeper networks struggled to propagate gradients effectively, leading to poor convergence with limited training data. Skip connections were introduced in the Residual Neural Network (ResNet) \cite{resnet} to play a significant role in mitigating this problem. It allowed gradients to flow directly through layers, to enable training of very deep networks. ResNet50, with 50 layers, offered an increased representation power of features. The ResNet backbones effectively extracted rich hierarchical features from the larger, well-curated datasets.

A contemporary architecture, InceptionNet \cite{inception}, took a step towards efficient deeper learning via smarter architectures. The multibranch architecture allowed the model to extract features at various levels of detail, making it useful for datasets with complex spatial structures. Inception modules reduced computational cost while preserving strong classification performance. This was advantageous for the screening of DR, which involved a variety of lesions of different sizes. However, the complex architecture demanded higher resolution input and underperformed on low-quality fundus images.
The Dense Convolutional Network (DenseNet) \cite{dense} improved gradient propagation and promoted feature reuse by employing dense connectivity. It facilitated efficient reuse of features while mitigating the problem of vanishing gradients. The DenseNet121 contained 121 layers. DenseNet gained popularity over DR datasets that have limited training data with a high diversity of lesions. However, it also resulted in increased memory usage and limited scalability for ultra-high-resolution images.

 As neural networks grew deeper, they achieved higher accuracy at the expense of increased data demand, computational complexity, and memory usage. This made them less suitable for real-time or resource-constrained applications. Consequently, there has emerged a growing demand for lightweight architectures to strike a balance between performance and efficiency. 
The EfficientNet \cite{efficient} was designed to further improve the accuracy-efficiency trade-off. It optimized the CNN architecture through a compound scaling strategy to simultaneously balance width, depth, and resolution. This led to a family of models, with EfficientNet-B2 offering an efficient trade-off between accuracy and computational cost; particularly, for high-dimensional images. It achieved state-of-the-art performance with fewer parameters compared to the other models. 

While CNNs relied on spatial locality and translation invariance, ViTs \cite{vit} treated an image as a sequence of non-overlapping patches, processing them with a self-attention mechanism. This enabled ViTs to model long-range dependencies and capture global relationships across an image, making them highly effective for complex vision tasks. They modeled long-range dependencies and contextual information more effectively than CNNs. However, ViTs required large datasets, such as EyePACS \cite{diabetic-retinopathy-detection}, for effective training; mainly due to the lack of built-in spatial priors. They struggled to generalize well in low-data scenarios, making them less effective than CNNs when training data was limited.

The CNN-based models mostly employ advanced data augmentation and pre-processing to address challenges like overfitting, limited training samples, and poor quality datasets. In many cases, models augmented with attention mechanisms or based on transformer architectures have the tendency to overfit. They do not consistently outperform well-designed, lower-complexity CNNs. This occurs particularly on smaller datasets or tasks that have limited variability. As a result, transfer learning with pre-trained CNN backbones remain a practical and effective strategy in data-constrained settings; particularly due to their lower computational demands and strong inductive biases for image-based tasks.

\subsection{Classification}
\label{subsec:classification}

This assigns an overall label to a fundus image based on (say) the severity grade of DR. It also includes binary categorization into RDR  and NRDR, as well as the identification of conditions such as DME. Clinically, the task is critical for the screening of large populations affected by diabetes and requires referral to an ophthalmologist. High sensitivity minimizes the risk of missing vision-threatening cases, while adequate specificity reduces unnecessary referrals. Severity grading guides urgency when prompt intervention is needed. The classification algorithms aim to match or surpass the accuracy of the expert graders in labeling DR stages from fundus images. They allow scalable early detection in primary care and telescreening programs. The task requires image-level class labels as ground-truth annotations.

Recent studies have focused on automating the diagnosis of DR to provide rapid, remote, and cost-effective diagnostic solutions. CNN-based architectures have been widely utilized for the diagnosis of DR, with consistently strong performance. Scarcity of training samples has prompted the adoption of transfer learning, involving pre-trained DL models. ResNet-50 \cite{resnet} backbone, pre-trained on the ImageNet \cite{deng2009imagenet}, was fine-tuned with the APTOS \cite{aptos2019} dataset incorporating data augmentation \cite{patil2023effective}, achieving 97.87\% accuracy for multiclass DR classification when evaluated in the EyePACs \cite{diabetic-retinopathy-detection} dataset. The pre-trained weights of ResNet-50, Inception-V3 \cite{inception}, VGG16 \cite{vgg}, DenseNet-121 \cite{dense}, MobileNet \cite{mobilenetv2}, and EfficientNetV2 \cite{efficient} were fine-tuned in an enhanced version of the EyePACs dataset for five-class categorization of DR \cite{incir2024study}, and VGG16 achieved the highest accuracy of 78.61\%. 

Although some CNNs were enhanced with attention mechanisms to improve contextual feature extraction, they often required high quality data for effective training. For example, ResNet-50 combined with a channel attention mechanism achieved an F1 score of 91.77\% for the three-class DME classification in Messidor \cite{fu2023automatic}. The dense connectivity integrated with attention mechanisms improved robustness, achieving 88.80\% accuracy on EyePACS for five-class DR classification \cite{cao2025diabetic}. 

Transformers have also been inducted for classification tasks. A transformer-guided Category-Relation Attention Network (CRA-Net) deduced pathological relationships among lesions and achieved classification accuracies of 89.1\% and 83.1\% in the APTOS and DDR datasets, respectively, for the five-class DR severity grading \cite{zang2024cra}. In many cases, models augmented with attention or based on transformers tended to overfit and did not consistently outperform well-designed, lower-complexity CNNs; particularly, on smaller datasets with limited variability. As a result, transfer learning with pre-trained CNN backbones remains a practical and effective strategy in data-constrained settings.

\subsection{Segmentation}
\label{subsec:segmentation}

It refers to the delineation of disease-related pathology, at the pixel level, in fundus images. Clinically, lesion segmentation offers a visual explanation and quantitative assessment of disease burden. For example, the number of microaneurysms and/or the area of exudates directly reflect the severity in the manifestation of DR. Identifying the fovea (as in Fig. \ref{fig:DRSeverity}), and measuring the distance of hard exudates from it, determines the presence of {DME \cite{rajalakshmi2025creating}. Segmentation highlights both the location and extent of the pathology, supporting interpretability and reliability beyond the overall grading. However, in practice, ophthalmologists rely mainly on image-level classification; with coarse localization of lesions offering additional support when needed. Fine-grained pixel-level segmentation, while valuable for research, often provides limited additional clinical utility. It requires detailed pixel-wise lesion annotations and is highly sensitive to the quality of fine-grained labels.

The $U$-Net was used for the semantic segmentation of red lesions \cite{saranya2023detection}. Advanced convolutional layers performed pixel-level labeling and the segmented output was fed to a custom CNN for binary DR classification, achieving 95.94\% accuracy on the IDRiD dataset. Incorporating attention improved lesion segmentation. A multiscale attention mechanism was integrated with FuDSA-Net to segment microaneurysms and hemorrhages in retinal images \cite{dey2023multi}. The CNN backbone extracted hierarchical features, while multiscale attention captured lesions across varying spatial resolutions. An F2-score of 0.6977 on Messidor and 0.6788 on IDRiD demonstrated its effectiveness. 

The Relation Transformer Network (RTNet) employed self-attention to capture global lesion dependencies and cross-attention to model interactions between lesion and vessel characteristics \cite{huang2022rtnet}. RTNet segmented four types of lesions, {\it viz.} hard exudates, hemorrhages, microaneurysms, and soft exudates, with varying AUC-PR scores (e.g., 90.24\% for hard exudates on IDRiD, but only 11.76\% for microaneurysms in DDR). These results highlight both the potential and limitations of transformer-based segmentation, particularly under constraints of dataset size and annotation quality.

\subsection{Detection}
\label{subsec:detection}

This encompasses the identification and localization of DR-related features or conditions within fundus images. Clinically, detection algorithms can support lesion counting, with fewer challenges than segmentation. They can help identify early signs that may get overlooked. Detection improves the granularity of screening systems while remaining closer to clinical needs. It requires coarse annotations, such as bounding boxes or approximate lesion markings.

CNN-based detection approaches have been widely explored. The Bounding Box Refinement Network was used for hemorrhage detection \cite{huang2020automated}, trained on 80 images from IDRiD and 590 private images. YOLO models, using CNN backbones, have been adopted to detect different pathologies in Messidor \cite{pal2020detection} and DDR \& IDRiD \cite{santos2022new}. Incorporating attention and asymmetric convolution branches improved robustness to rotated or flipped images, enabling more accurate lesion detection in DR \cite{cao2025diabetic} on the EyePACS dataset. Despite their complexity, attention-based and transformer-enhanced detection models have not consistently surpassed well-optimized CNN-based detectors; particularly, in smaller datasets.

\section{Data Repositories} 
\label{sec:background}

The growing need for regular screening for DR in the population affected by diabetes required a continuous effort in the development of automated DL-based screening tools to help in its early diagnosis. These are expected to be beneficial, especially in remote and underprivileged regions. As the success of deep networks is closely related to the quality and quantity of data, research in this domain necessitated the creation of publicly available datasets. This section describes some of the well-known open access fundus image datasets related to DR, after highlighting some of the existing challenges.

\subsection{Challenges in databases}
\label{sec:challenges}

Developing high-quality, robust, and standardized datasets for DR screening poses challenges in their acquisition, annotation, and curation. Fig. \ref{fig:challenge} illustrates some of the challenges, grouping them into the major stages of the dataset development process.

\begin{figure*}[t]
\centerline{\includegraphics[width=\textwidth]{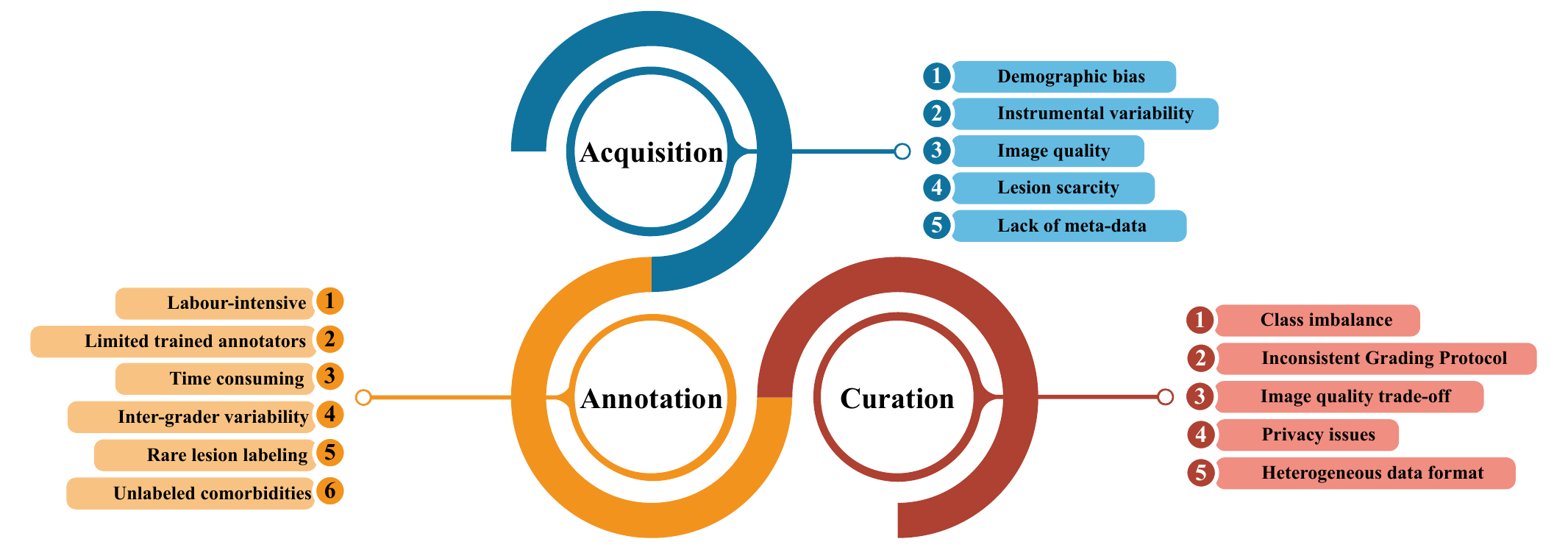}}
\caption{Key challenges in developing high-quality, robust and standardized datasets for DR screening.}
\label{fig:challenge}
\end{figure*}

\noindent {\textbf{Acquisition:} Gathering images from diverse populations is difficult, often leading to demographic bias. Differences in the type of camera used, its resolution, and field of view induce variability. Poor illumination, blur, and prevalence of other artifacts frequently result in low-quality images. Many datasets also lack consistent acquisition protocols and/or rich metadata; thereby, limiting their value.

\noindent {\textbf{Annotation:} Accurate annotation or labelling depends on expert graders. However, this process is labor intensive, time consuming, and expensive. In a highly populated country like India, the ratio between the ophthalmologist and the population ranges from 1:9,000 to as low as 1:600,000 in certain regions \cite{kumar2021diabetic}. This restricts regular large-scale manual screening, making it impractical.  The absence of standardized use of annotation guidelines, such as the Early Treatment Diabetic Retinopathy Study (ETDRS) \cite{ETDRS} or the International Clinical Diabetic Retinopathy (ICDR) severity scale \cite{wilkinson2003proposed}, in each collection center, results in inter-grader variability and label inconsistency. Fine-grained annotation of the lesion is particularly difficult to obtain, despite its importance in understanding the different stages of DR. Certain lesions may occur infrequently, leading to under-representation in datasets. Images may simultaneously contain other ocular diseases, such as cataracts, glaucoma, or Age-related Macular Degeneration (AMD), while lacking multiple labeling. 

\noindent {\textbf{Curation:} This refers to the systematic process of selecting, cleaning, annotating, and organizing images, to build a reliable dataset for developing a model. Class imbalance, inconsistent quantification of the DR severity measure, heterogeneous data formats, and variable annotation schemes complicate dataset curation. Privacy and data-sharing regulations and/or constraints restrict the availability of datasets. Although high-quality images are needed for training, their real-time clinical deployment often involves lower-quality inputs. This requires a trade-off in determining which images to retain during curation. Datasets collected across multiple centers frequently differ in resolution, format, and annotation protocol. This hinders their integration.

\begin{table}[!t]
\caption{Key characteristics of DR datasets and their impact.}
\label{tab:characteristics}
\centering
\setlength{\tabcolsep}{6pt}
\renewcommand{\arraystretch}{1.25}

\begin{tabular}{p{0.22\textwidth} p{0.32\textwidth} p{0.32\textwidth}}
\hline
\rowcolor{RowHeader}
\textbf{Characteristic} & \textbf{Suitability for DL} & \textbf{Clinical Implication} \\
\hline

Large and Diverse
& \begin{tabular}{l}
    - Enhances generalization across\\populations and imaging devices.
\end{tabular} 
& \begin{tabular}{l}
     - Enables robust performance across\\ diverse populations and settings.
\end{tabular}\\
\hline

\rowcolor{RowAlt}
Well-Annotated
& - Enables effective supervised training.
& \begin{tabular}{l}
     - Supports accurate diagnosis and\\ lesion-level decision-making.
\end{tabular}\\
\hline

Standardized Protocols
& \begin{tabular}{l}
     - Provides consistent ground truth\\ and fair benchmarking.
\end{tabular}
& - Improves clinical reliability. \\
\hline

\rowcolor{RowAlt}
Balanced Representation
& \begin{tabular}{l}
     - Reduces bias toward majority classes;\\- Improves learning of rare lesions.
\end{tabular}
& \begin{tabular}{l}
     - Improves sensitivity for rare\\but severe cases.
\end{tabular}\\
\hline

Rich Metadata
& \begin{tabular}{l}
     - Supports multi-modal models;\\- Disentangles DR-specific effects\\from other factors.
\end{tabular}
& \begin{tabular}{l}
     - Adds context for prognosis and\\risk assessment.
\end{tabular}\\
\hline

\rowcolor{RowAlt}
Longitudinal Data
& \begin{tabular}{l}
     - Enables progression modeling and\\predictive analytics.
\end{tabular}
& \begin{tabular}{l}
     - Allows monitoring of disease\\evolution;\\- Supports prediction of treatment\\response and disease control.
\end{tabular}\\
\hline

Clinical Utility Features
& - Aligns models with real screening tasks.
& \begin{tabular}{l}
     - Supports early identification\\and prioritization of high-risk\\cases. 
\end{tabular}\\
\hline
\end{tabular}
\end{table}

\begin{figure*}[t]
\centerline{\includegraphics[width=\textwidth]{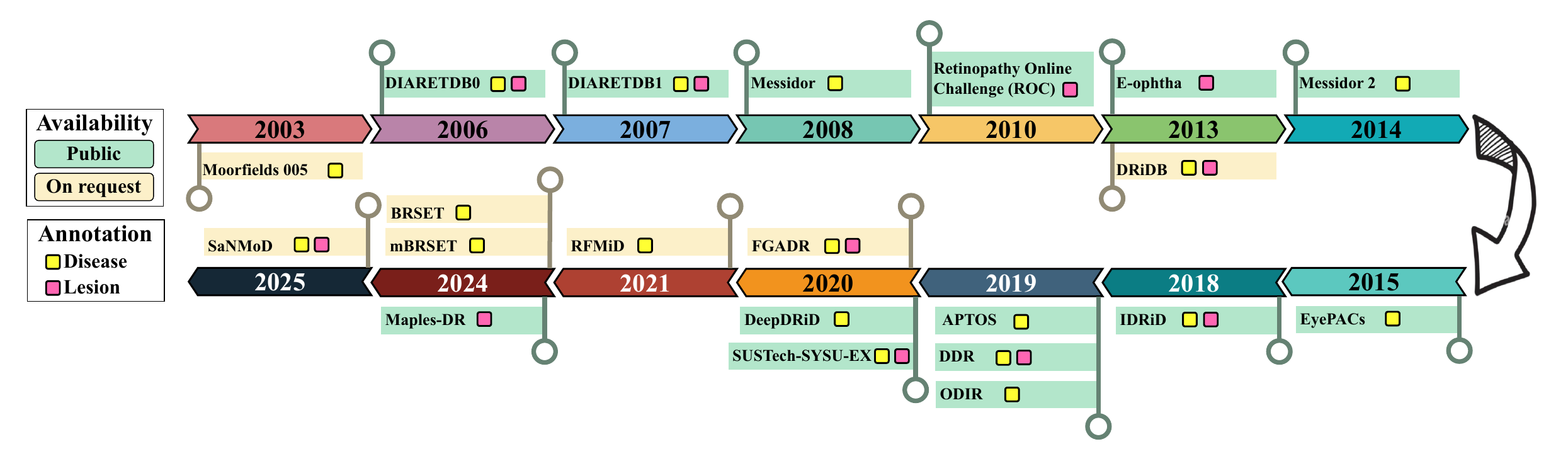}}
\caption{Chronological overview of major DR fundus image datasets (from 2003 to 2025), showing year of release, dataset type (image-level vs. lesion annotation), and availability.}
\label{fig:dataset_timeline}
\end{figure*}

The challenges mentioned above highlight some of the gaps that affect the design and utility of DR datasets. Automated DL systems, trained on well-curated datasets, have demonstrated the potential to lower diagnostic costs by improving accessibility and precision. They also provide valuable support to healthcare facilities, with performance approaching expert-level accuracy. Table~\ref{tab:characteristics} summarizes key characteristics for determining the suitability of a dataset for DL applications and their clinical use.

\subsection{Available datasets}
\label{subsec:datasets}

Several fundus image repositories, including Messidor \cite{messidor}, EyePACS \cite{diabetic-retinopathy-detection}, APTOS \cite{aptos2019}, IDRiD \cite{idrid}, DDR \cite{li2019diagnostic}, E-ophtha \cite{decenciere2013teleophta}, and SanMoD \cite{rajalakshmi2025creating}, have been developed in recent years to advance research in the field of DR. Most of these datasets originate from limited demographic regions. This restricts their ability to capture the full spectrum of DR manifestations across diverse cohorts of patients, as necessary for their real-world deployment. Each dataset carries inherent limitations, such as restricted size, inconsistent annotation quality, or incomplete labeling. Even within India \cite{idrid}, several datasets suffer from small sample sizes or inadequate grading information; thereby, reducing their suitability to build robust and generalizable DR detection models tailored to the local population. The limitations underscore the importance of tracing how the DR datasets evolved over time.

Fig. \ref{fig:dataset_timeline} illustrates the development and availability of DR datasets over the years. In the past decade, the growing demand for automated DR diagnosis drove the release of multiple repositories. Most of these datasets provide disease-level annotations, as image-level labeling is less expensive and/or labor intensive. However, only a few offer annotations at the lesion level curated by trained specialists to advance studies focused on lesions. Alongside the publicly accessible resources, several datasets remained private; with some made available on request, under well-defined research agreements. Many studies also used private datasets. This study focuses on datasets that are publicly available or accessible on request. Together, such datasets support progress ranging from broad disease classification to fine-grained lesion analysis. They provide the essential foundation for clinical development of artificial intelligence (AI).

These fundus image datasets can broadly be divided into two generations. Table \ref{tab:pubDatasetStat_2014} presents the earlier datasets developed between 2003 and 2014. Researchers created these relatively small, lesion-centric datasets mainly to establish benchmark problems for lesion detection and coarse DR grading. They served as proof-of-concept resources for algorithm development, while remaining limited in scale, demographic diversity, and metadata. Most annotations covered only MA, HE, HX, and SX, while crucial advanced-stage lesions such as IRMA or NV were absent. The first generation of datasets established early benchmarks but were small and lesion-centric.

The limitations of first generation datasets prompted a clear shift toward large-scale, multi-tasking, clinically-oriented resources. With the availability of faster internet and cheaper hardware resources, this became a reality.  Table \ref{tab:pubDatasetStat_2025} summarizes the recent datasets, released between 2015 and 2025, encompassing tens of thousands of images from multiple centers and diverse populations. They often provide rich metadata, multiple expert graders, and annotations that span DR along with other ocular diseases. A stronger association of DR is captured with other ocular diseases. Developers thus expanded the lesion annotations, to include a broader spectrum of DR-related anomalies, moving beyond the earlier focus on the optic disc and vessel markings. The evolution marked a transition from small, curated benchmarks for methodological testing, to expansive, community-driven datasets that supported DL at larger scale. This enabled generalization across cohorts, with advanced clinical deployment.

\begin{table}[!t]
\caption{Details of DR datasets developed during 2003 - 2014.}
\label{tab:pubDatasetStat_2014}
\begingroup

\setlength{\tabcolsep}{6pt} 
\renewcommand{\arraystretch}{1.25} 
\scalebox{0.65}{
\begin{tabular}{llrcccccl}
\hline
\textbf{Dataset}                                                                                               & \multicolumn{1}{c}{\textbf{Origin}} & \multicolumn{1}{c}{\textbf{\# Samples}}                  & \textbf{\# Patients}       & \textbf{Resolution}                                                                                 & \textbf{\begin{tabular}[c]{@{}c@{}}DR \\ Grade\end{tabular}} & \textbf{\begin{tabular}[c]{@{}c@{}}Lesion \\ Types;\\ Annotation\\ (Coarse/ Fine)\end{tabular}} & \textbf{\#Annotators} & \multicolumn{1}{c}{\textbf{Comment}}                                                                                                                                                                                                                                                                            \\ \hline
\begin{tabular}[c]{@{}l@{}}\textbf{Moorfields 005} \\ \cite{moorfields005}\end{tabular}                                   & UK                                  & 5,537,798                                                & \multicolumn{1}{r}{45,824} & Unknown                                                                                             & 5                                                            & NA                                                                                              & -                     & \begin{tabular}[c]{@{}l@{}}Longitudinal data, Available meta data,\\ (Reported as of July 2024)\end{tabular}                                                                                                                                                                                                    \\ \hline
\begin{tabular}[c]{@{}l@{}}\textbf{DIARETDB0} \\ \cite{kauppi2006diaretdb0}\end{tabular}                                  & Finland                             & 130                                                      & -                          & 1500 $\times$ 1152                                                                                  & 5                                                            & \begin{tabular}[c]{@{}c@{}}MA, HE, \\ HX, SX;\\ Coarse\end{tabular}                             & 4                     & \begin{tabular}[c]{@{}l@{}}Lesions are marked with circle, ellipse,\\ oval shapes or rarely polygon\end{tabular}                                                                                                                                                                                                \\ \hline
\begin{tabular}[c]{@{}l@{}}\textbf{DIARETDB1} \\ \cite{kalviainen2007diaretdb1}\end{tabular}                              & Finland                             & 89                                                       & -                          & 1500 $\times$ 1152                                                                                  & 5                                                            & \begin{tabular}[c]{@{}c@{}}MA, HE,\\ HX, SX;\\ Coarse\end{tabular}                              & 4                     & \begin{tabular}[c]{@{}l@{}}Lesions are marked with circle, ellipse,\\ oval shapes or rarely polygon\end{tabular}                                                                                                                                                                                                \\ \hline
\begin{tabular}[c]{@{}l@{}}\textbf{Messidor} \\ \cite{messidor}\end{tabular}                                              & France                              & 1,200                                                    & -                          & \begin{tabular}[c]{@{}c@{}}1440 $\times$ 960,\\ 2241$\times$1488,\\ 2304 $\times$ 1536\end{tabular} & 4                                                            & NA                                                                                              & -                     & Contains grade risk of DME (0, 1, 2)                                                                                                                                                                                                                                                                            \\ \hline
\begin{tabular}[c]{@{}l@{}}\textbf{Retinopathy Online} \\ \textbf{Challenge (ROC)} \\ \cite{niemeijer2009retinopathy}\end{tabular} & USA                                 & 100                                                      & -                          & \begin{tabular}[c]{@{}c@{}}768 $\times$ 576,\\ 1058$\times$1061,\\ 1389 $\times$ 1383\end{tabular}  & -                                                            & MA                                                                                              & 4                     & \begin{tabular}[c]{@{}l@{}}Annotations include MA only,\\ The center location of MA are marked,\\ Lesion structurally similar to MA also marked\end{tabular}                                                                                                                                                    \\ \hline
\begin{tabular}[c]{@{}l@{}}\textbf{E-ophtha} \\ \cite{decenciere2013teleophta}\end{tabular}                               & France                              & \begin{tabular}[c]{@{}r@{}}EX: 82\\ MA: 351\end{tabular} & -                          & \begin{tabular}[c]{@{}c@{}}1440 $\times$ 960 to\\ 2544 $\times$ 1696\end{tabular}                   & NA                                                           & \begin{tabular}[c]{@{}c@{}}MA, HX,\\ SX;\\ Fine\end{tabular}                                    & -                     & \begin{tabular}[c]{@{}l@{}}E-ophtha is divided into two set: \\ E-ophtha-MA, E-ophtha-EX,\\ E-ophtha-MA contains MA annotation,\\ E-ophtha-EX contains exudate (EX) annotation,\\ Multiple images per patient is available,\\ Overlap of samples between E-ophtha-MA and \\ E-ophtha-EX is unknown\end{tabular} \\ \hline
\begin{tabular}[c]{@{}l@{}}\textbf{DRiDB} \\ \cite{prentavsic2013diabetic}\end{tabular}                                   & Croatia                             & 50                                                       & -                          & 720 $\times$ 576                                                                                    &                                                              & \begin{tabular}[c]{@{}c@{}}MA, HE,\\ HX, SX;\\ Coarse\end{tabular}                              & 5                     & \begin{tabular}[c]{@{}l@{}}Blood Vessel, Optic Disc and Macula also annotated,\\ NV annotation\end{tabular}                                                                                                                                                                                                     \\ \hline
\begin{tabular}[c]{@{}l@{}}\textbf{Messidor-2} \\ \cite{messidor}\end{tabular}                                            & France                              & 1,748                                                    & -                          & \begin{tabular}[c]{@{}c@{}}1440 $\times$ 960,\\ 2241$\times$1488,\\ 2304 $\times$ 1536\end{tabular} & -                                                            & NA                                                                                              & -                     & \begin{tabular}[c]{@{}l@{}}Macula-centered images,\\ Information about left/ right eye,\\ Extension of Messidor dataset,\\ Only third party annotations available\end{tabular}                                                                                                                                  \\ \hline
\end{tabular}}
\endgroup%
\end{table}

\begin{table}[!t]
\caption{Details of DR datasets developed during 2015 - 2025.}
\label{tab:pubDatasetStat_2025}
\begingroup

\setlength{\tabcolsep}{6pt} 
\renewcommand{\arraystretch}{1.25} 
\scalebox{0.65}{
\begin{tabular}{llrcccccl}
\hline
\textbf{Dataset}                                                                                  & \multicolumn{1}{c}{\textbf{Origin}} & \multicolumn{1}{c}{\textbf{\# Samples}} & \textbf{\# Patients}      & \textbf{Resolution}                                                                                    & \textbf{\begin{tabular}[c]{@{}c@{}}DR \\ Grade\end{tabular}} & \textbf{\begin{tabular}[c]{@{}c@{}}Lesion \\ Types;\\ Annotation\\ (Coarse/ Fine)\end{tabular}}                                                                & \textbf{Annotators} & \multicolumn{1}{c}{\textbf{Comment}}                                                                                                                                                                                                                                                                                                              \\ \hline
\begin{tabular}[c]{@{}l@{}}\textbf{EyePACS} \\ \cite{diabetic-retinopathy-detection}\end{tabular} & USA                                 & 88,702                                  & -                         & \begin{tabular}[c]{@{}c@{}}433 $\times$ 289 to \\ 5184 $\times$ 3456\end{tabular}                      & 5                                                            & NA                                                                                                                                                             & -                   & \begin{tabular}[c]{@{}l@{}}Both left and right eye images of \\ each patient provided\end{tabular}                                                                                                                                                                                                                                                \\ \hline
\begin{tabular}[c]{@{}l@{}}\textbf{IDRiD} \\ \cite{idrid}\end{tabular}                            & India                               & 516                                     & -                         & 4288 $\times$ 2848                                                                                     & 5                                                            & \begin{tabular}[c]{@{}c@{}}MA, HE,\\ HX, SX;\\ Fine\end{tabular}                                                                                               & -                   & \begin{tabular}[c]{@{}l@{}}81 images have fine lesion annotation,\\ 516 images have classification labels with \\ coarse annotation for fovea and optic disc,\\ Contains DME grade (0, 1)\end{tabular}                                                                                                                                            \\ \hline
\begin{tabular}[c]{@{}l@{}}\textbf{APTOS} \\ \cite{aptos2019}\end{tabular}                        & India                               & 5,590                                   & -                         & \begin{tabular}[c]{@{}c@{}}474 $\times$ 358 to\\ 4288 $\times$ 2848\end{tabular}                       & 5                                                            & NA                                                                                                                                                             & -                   & \begin{tabular}[c]{@{}l@{}}Macula centered images,\\ Only 3,662 samples are publicly available\end{tabular}                                                                                                                                                                                                                                       \\ \hline
\begin{tabular}[c]{@{}l@{}}\textbf{DDR} \\ \cite{li2019diagnostic}\end{tabular}                   & China                               & 13,673                                  & \multicolumn{1}{r}{9,598} & \begin{tabular}[c]{@{}c@{}}702 $\times$ 706 to\\ 5184 $\times$ 3456\end{tabular}                       & 5                                                            & \begin{tabular}[c]{@{}c@{}}MA, HE,\\ HX, SX;\\ Coarse\end{tabular}                                                                                             & 7                   & Very high variability in image resolution                                                                                                                                                                                                                                                                                                         \\ \hline
\begin{tabular}[c]{@{}l@{}}\textbf{ODIR} \\ \cite{li2020benchmark}\end{tabular}                   & China                               & 10,000                                  & \multicolumn{1}{r}{5,000} & 2240 $\times$ 1488                                                                                     & 2                                                            & NA                                                                                                                                                             & -                   & \begin{tabular}[c]{@{}l@{}}Each image categorized to 8 ocular diseases\\ classes: Normal, DR, Glaucoma, \\ Age-related macular degeneration, Cataract, \\ Hyper-tension, Pathological Myopia, Others,\\ Meta-data about gender, age, etc.,\\ Images of left and right eye for each patient\end{tabular}                                           \\ \hline
\begin{tabular}[c]{@{}l@{}}\textbf{DeepDRiD} \\ \cite{LIU2022100512}\end{tabular}                 & China                               & 2,128                                   & \multicolumn{1}{r}{756}   & \begin{tabular}[c]{@{}c@{}}1956 $\times$ 1934,\\ 3900 $\times$ 3072\end{tabular}                       & 5                                                            & NA                                                                                                                                                             & -                   & \begin{tabular}[c]{@{}l@{}}Dual-view images of each eye - optic disc- \\ and fovea-centered,\\ Each image has grades for quality,\\ DR has grades from 0 to 5; 0 - Normal,\\ 1 - Mild, 2 - Moderate, 3 - Severe,\\ 4 - PDR, 5 - Ungradable,\\ Patient level DR grade,\\ Only 1,702 samples are publicly available,\\ Other Meta data\end{tabular} \\ \hline
\begin{tabular}[c]{@{}l@{}}\textbf{SUSTech-SYSU-EX} \\ \cite{lin2020sustech}\end{tabular}         & China                               & 1,219                                   & -                         & 2880 $\times$ 2136                                                                                     & 5                                                            & \begin{tabular}[c]{@{}c@{}}HX, SX;\\ Coarse\end{tabular}                                                                                                       & 3                   & \begin{tabular}[c]{@{}l@{}}Only exudate annotation,\\ Contains location of optic disc \& fovea\end{tabular}                                                                                                                                                                                                                                       \\ \hline
\begin{tabular}[c]{@{}l@{}}\textbf{FGADR} \\ \cite{zhou2021benchmark}\end{tabular}                & UAE                                 & 2,842                                   & -                         & Unknown                                                                                                & 5                                                            & \begin{tabular}[c]{@{}c@{}}MA, HE, HX,\\ SX, IRMA, NV,\\ Laser mark, Proliferate,\\ Membrane;\\ Fine\end{tabular}                                              & 6                   & \begin{tabular}[c]{@{}l@{}}All images has DR grade,\\ 1,842 images has lesion annotation\end{tabular}                                                                                                                                                                                                                                             \\ \hline
\begin{tabular}[c]{@{}l@{}}\textbf{RFMiD} \\ \cite{pachade2021retinal}\end{tabular}               & India                               & 3,200                                   & -                         & \begin{tabular}[c]{@{}c@{}}2144 $\times$ 1424,\\ 4288 $\times$ 2848,\\ 2048 $\times$ 1536\end{tabular} & 2                                                            & NA                                                                                                                                                             & 2                   & \begin{tabular}[c]{@{}l@{}}Images have 28 different disease label\\ with 46 diseases,\\ Marks only presence or absence of disease,\\ Contains DME grade (0, 1)\end{tabular}                                                                                                                                                                       \\ \hline
\begin{tabular}[c]{@{}l@{}}\textbf{MAPLES-DR} \\ \cite{maples_dr}\end{tabular}                    & France                              & 198                                     & -                         & 1500 $\times$ 1500                                                                                     & -                                                            & \begin{tabular}[c]{@{}c@{}}MA, HE, HX,\\ SX, NV, Drusen,\\ Vessel, Macula,\\ Optic Cup \& Disc;\\ Fine\end{tabular}                                            & 7                   & \begin{tabular}[c]{@{}l@{}}Images belong to Messidor \cite{messidor}, \\ For uniformity, images in Messidor are resized\end{tabular}                                                                                                                                                                                                              \\ \hline
\begin{tabular}[c]{@{}l@{}}\textbf{BRSET} \\ \cite{nakayama2023brazilian}\end{tabular}            & Brazil                              & 16,266                                  & \multicolumn{1}{r}{8,524} & \begin{tabular}[c]{@{}c@{}}951 $\times$ 874 to\\ 2984 $\times$ 2304\end{tabular}                       & 5                                                            & NA                                                                                                                                                             & 1                   & \begin{tabular}[c]{@{}l@{}}Macula centered images,\\ Contains DME grade (0, 1),\\ Meta data include nationality, age, sex, clinical\\ antecedents, insulin use, and diabetes time\end{tabular}                                                                                                                                                    \\ \hline
\begin{tabular}[c]{@{}l@{}}\textbf{mBRSET} \\ \cite{nakayama2024mbrset}\end{tabular}              & Brazil                              & 5,164                                   & \multicolumn{1}{r}{1,291} & 1600 $\times$ 1600                                                                                     & 5                                                            & NA                                                                                                                                                             & 3                   & \begin{tabular}[c]{@{}l@{}}Image are captured using Handheld smartphone-\\ based fundus camera,\\ Contains DME grade (0, 1),\\ Include both macula centered and optic disc \\ centered images\end{tabular}                                                                                                                                        \\ \hline
\begin{tabular}[c]{@{}l@{}}\textbf{SaNMoD} \\ \cite{rajalakshmi2025creating}\end{tabular}         & India                               & 4,212                                   & -                         & \begin{tabular}[c]{@{}c@{}}1504 $\times$ 1000,\\ 1920 $\times$ 991,\\ 2588 $\times$ 1958\end{tabular}  & 5                                                            & \begin{tabular}[c]{@{}c@{}}MA, HE, HX, SX,\\ IRMA, Venous,\\ Looping, Laser Mark,\\ Fibrous, Proliferans,\\ NV on Disc, \\ NV elsewhere;\\ Coarse\end{tabular} & 8                   & \begin{tabular}[c]{@{}l@{}}Macula centered images, \\ Contains labels (0, 1) of RDR and DME, \\ 11 different DR related lesions are marked,\\ Gradable/ un-gradable image labels\end{tabular}                                                                                                                                                     \\ \hline
\end{tabular}}
\endgroup%
\end{table}

All datasets in Tables \ref{tab:pubDatasetStat_2014} and \ref{tab:pubDatasetStat_2025}, having DR grades, support binary or multi-class grading. However, several challenges hinder their use in DL. Class imbalance persists across grades, large-scale models demand extensive data, and supervised training remains highly sensitive to label noise. Population-specific anatomical variation further limits generalization, as models trained on data specific to one region (say, the USA) often under-perform on another (say, Asia). Large-scale datasets, such as EyePACS \cite{diabetic-retinopathy-detection} and APTOS \cite{aptos2019}, suffer from noisy labels, while DDR \cite{li2019diagnostic} faces high variability in image resolution. Although many China-based datasets \cite{li2019diagnostic, LIU2022100512, lin2020sustech} provide scale, they remain demographically biased. Over the past decade, India has produced large-scale datasets such as IDRiD \cite{idrid}, APTOS \cite{aptos2019}, and RFMiD \cite{pachade2021retinal}. The most recent addition, SaNMoD \cite{rajalakshmi2025creating}, provides high-resolution images, a large sample size, and labels from multiple annotators. This increases its reliability. Analogously, Brazil has contributed the datasets BRSET and mBRSET \cite{nakayama2023brazilian, nakayama2024mbrset}. The emergence of large-scale datasets from India, China, the USA, and Brazil, reflects a global trend toward generating population-specific resources to mitigate bias and enhance generalizability.

Although segmentation has limited direct clinical significance, it contributes to the understanding of lesions associated with DR and supports the interpretability of models. Any dataset with lesion annotations can serve a segmentation task. However, coarse annotations can often include healthy pixels and lead to ambiguous supervision. Fine lesion annotations can provide the most relevant supervision for a segmentation. As shown in Tables \ref{tab:pubDatasetStat_2014} and \ref{tab:pubDatasetStat_2025}, very few datasets contain fine lesion annotations. Those that do possess fine annotations have only a small number of samples, making segmentation challenging due to data scarcity. Early datasets \cite{decenciere2013teleophta, prentavsic2013diabetic} primarily annotated preliminary DR lesions such as MA, HE, HX, and SX, while some \cite{niemeijer2009retinopathy} focused on a single lesion type. Recent datasets such as FGADR \cite{zhou2021benchmark} and MAPLES-DR \cite{maples_dr} expanded lesion categories, reflecting growing interest in lesion-level understanding of DR for improved interpretability of DL models. High-resolution images remain essential for segmentation, as clear visibility of pathological features is critical for an accurate performance of automated systems.

Any dataset containing coarse or fine lesion annotations can serve for lesion localization tasks. Coarse annotations, such as circles or ellipses in DIARETDB0/1 \cite{kauppi2006diaretdb0, kalviainen2007diaretdb1} and DRiDB \cite{prentavsic2013diabetic}, allow only approximate localization. Annotations in datasets such as E-ophtha \cite{decenciere2013teleophta}, IDRiD \cite{idrid}, FGADR \cite{zhou2021benchmark}, DDR \cite{li2019diagnostic}, and SaNMoD \cite{rajalakshmi2025creating}, provide more reliable lesion-level localization. The annotation of 11 different pathologies in SaNMoD dataset helps to capture diverse pathologies relevant to advanced stages of DR. Some datasets, including the ROC challenge \cite{niemeijer2009retinopathy} and SUSTech-SYSU-EX \cite{lin2020sustech}, concentrate on a single lesion type. This creates valuable but narrow benchmarks. The involvement of multiple expert annotators in DDR and SaNMoD enhances annotation reliability. Among all, SaNMoD offers the largest number of samples with the most comprehensive DR-related lesion annotations. Its consistent, high-resolution images make it particularly suitable for DR screening in the Indian population. Overall, the lesion detection datasets support training models to localize early signs of DR. This is a task with strong clinical utility, since lesion location and extent directly influence staging of the disease and its treatment planning.

Although many studies have addressed the automated screening of DR, the disease often co-occurs with glaucoma, age-related macular degeneration, and some other ocular disorders. Datasets labeled only for DR frequently contain images affected by additional pathologies. This reduces reliability in DR screening. Multi-disease datasets such as ODIR \cite{li2020benchmark} and RFMiD \cite{pachade2021retinal} provide richer annotations for improved model robustness. Messidor \cite{messidor}, IDRiD \cite{idrid}, BRSET \cite{nakayama2023brazilian}, mBRSET \cite{nakayama2024mbrset}, and SaNMoD \cite{rajalakshmi2025creating} additionally include labels for DME, a critical factor for risk of blindness. Hence these datasets help in extending DR detection to a multi-disease diagnosis scenario, and enhance the clinical relevance of the automated system.
Diabetes, combined with factors such as age and hypertension, strongly influence the development and progression of DR. Metadata capturing such information, as provided in Moorfields 005 \cite{moorfields005}, ODIR \cite{li2020benchmark}, DeepDRiD \cite{LIU2022100512}, and BRSET \cite{nakayama2023brazilian}, can significantly enhance the value of the dataset. Medication history, glycemic control, smoking, diet, and exercise also serve as important prognostic modifiers of DR. Tracking such information in future datasets can improve the clinical utility of automated systems.

DR is a progressive irreversible disorder that worsens without treatment. However, the rate of progression of the disease can be minimized with early detection through effective screening. The condition therefore requires continuous monitoring over time. Datasets that compile longitudinal images of DR patients, such as Moorfields 005 \cite{moorfields005}, are thus valuable to guide treatment strategies and support clinical decision making.

\section{A Benchmarking Study}
\label{sec:sanmod}

This section analyzes the recently published SaNMoD \cite{rajalakshmi2025creating} dataset, highlighting some key characteristics of DR datasets that influence model selection and performance. 
It is one of the most recent large-scale fundus image datasets, comprising more than 4,200 high-resolution samples captured with modern imaging devices. Annotated by eight ophthalmologists, it provides labels for DR stages, DME, and lesion-level features; thus contributing as a rich resource for database. These characteristics position SaNMoD as a representative of the current generation of datasets, closely aligned with clinical workflows, and supporting both disease-and lesion-level annotations for automated screening.

\begin{figure}[htbp]
\centering
\includegraphics[width=1.0\textwidth]{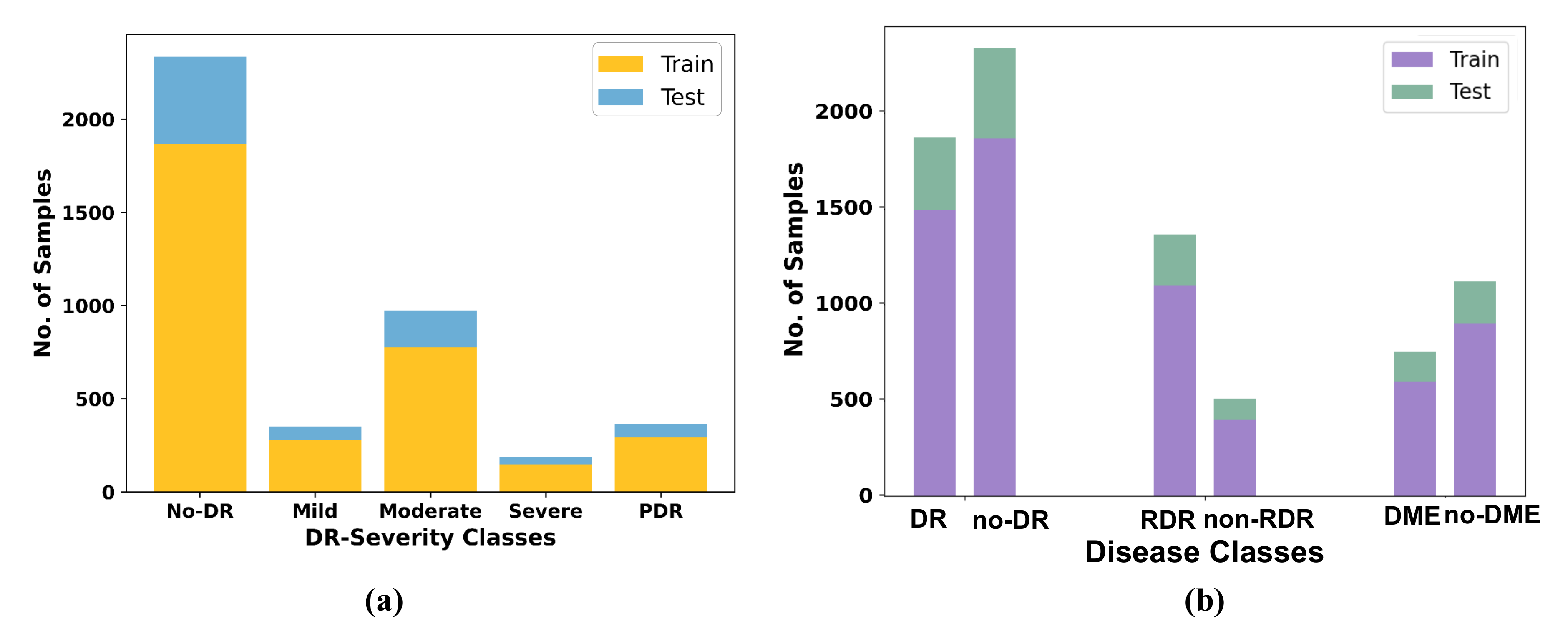}
\caption{Data distribution in SaNMoD, in terms of (a) the DR severity classes (grades), and (b) samples in the different categories.}
\label{fig:DataDist}
\end{figure}

\subsection{Data description and pre-processing}

The SaNMoD dataset contains 4,212 samples and exhibits a significant class imbalance, {\it viz.} 2,339 normal, 350 mild, 972 moderate, 187 severe, and 364 cases of PDR. This imbalance reflects real-world screening populations, where most patients present without DR or with a moderate level of the disease. Fig.~\ref{fig:DataDist} illustrates the data distribution of SaNMoD with reference to the different classes and tasks. The inclusion of DME labels, as well as referable DR (RDR) from severity grades, enhance its clinical relevance; thereby, making SaNMoD particularly suitable in evaluating binary detection, referral-level classification, and multi-class severity grading.
Four clinically relevant classification tasks were considered to assess the SaNMoD dataset. These are binary DR classification (No-DR vs. DR), referable DR classification (Non-RDR vs. RDR), DME classification (No-DME vs. DME), and multi-class DR severity grading (No-DR, Mild, Moderate, Severe, PDR). The tasks mirror the real-world diagnostic and screening workflows, where disease detection, referral decision, and severity grading directly influence treatment planning. In the case study reported here, the dataset was evaluated solely on classification tasks.
It was pre-processed to ensure compatibility with DL architectures. Only images in standard RGB format were retained, while corrupted or non-conforming files were excluded. This curation step removed two images and produced a set of 4,200 usable samples suitable for training automated diagnostic models.

As DL models require fixed input dimensions, image standardization becomes essential for training. Direct use of irregular images leads to structural conflicts, while square resizing distorts the natural aspect ratio and compromises spatial features. To preserve fidelity, a rectangular resizing strategy was applied. The inputs were set at $512\times340$ for CNN-based models and $224\times224$ for ViT. No additional preprocessing, such as augmentation or denoising, was introduced to maintain benchmark integrity.
Pre-trained CNN and ViT models were implemented in PyTorch using the Torchvision library to evaluate the compatibility of DL architectures with SaNMoD. Training was carried out on a single NVIDIA 12GB GPU for 150 epochs, with a batch size of 8 and a learning rate of 0.0001. All models were initialized using ImageNet weights, with parameters updated by AdamW optimizer \cite{loshchilovdecoupled} and a weight decay of 0.0001. The output layers were modified to match the number of target classes, followed by sigmoid activation for binary classification and softmax activation for multi-class grading. The model with the lowest validation loss was selected to enhance generalization in the test split. Binary cross-entropy loss with class weights was used as the objective function for binary classification. Weighted cross-entropy, based on the DR severity distribution of SaNMoD, was applied for multi-class grading. These strategies penalized the misclassification of minority classes, such as Mild, Severe, and PDR, and improved robustness while maintaining clinical relevance.

\subsection{Performance analysis}
\label{subsec:permformance_analysis}

The performance of multiple DL architectures was analyzed on the SaNMoD dataset, in clinically relevant tasks. The evaluation highlights how the characteristics of the dataset, such as lesion subtlety, class imbalance, and overlapping severity grades, influence the model results. The results provide a benchmark for assessing the utility of the dataset, and reveal both its strengths and limitations in supporting automated DR screening.
A suite of CNN and transformer models, including VGG16, ResNet50, InceptionNetV3, DenseNet121, EfficientNetB2 and ViT, was used to benchmark the SaNMoD dataset. CNNs leveraged spatial inductive biases that effectively captured localized lesions such as microaneurysms and hemorrhages. While the ViTs attempted to model long-range dependencies, they showed limited performance in this data-scarce, imbalanced setting. The results are based on common characteristics of the DR datasets, like  subtle morphology of the lesion, variable image quality, and skewed class distributions, and consistently shaped the performance of the model. Weighted binary cross-entropy and weighted cross-entropy losses aligned with these challenges, penalizing misclassification of minority classes. In general, CNNs provided stronger and more reliable baselines. The transformer-based models underperformed, underscoring the impact of dataset imbalance and lesion-level complexity on model robustness.

The overall performance of the models on SaNMoD was evaluated using Accuracy (Acc.), Balanced Accuracy (B.Acc.), F1-score,  and Area Under the Precision-Recall curve (AUC-PR) \cite{pedregosa2011scikit}. The AUC-PR served as the primary metric, as it provides a more reliable assessment in imbalanced datasets by emphasizing the precision-recall trade-off for minority classes. This choice is particularly relevant for DR datasets, here SaNMoD, where Mild and PDR cases remain under-represented. 

Table \ref{tab:DR_NoDR} summarizes the performance of DL models with the SaNMoD dataset, for binary classification of DR. The data provided sufficient discriminative information for automated detection of DR. All CNN-based models achieved F1-scores close to 0.90 for both DR and No-DR classes, indicating their ability to capture disease-relevant features such as microaneurysms and hemorrhages. EfficientNetB2 achieved the strongest performance across all metrics, suggesting its robustness and generalization potential. In contrast, the ViT lagged significantly, highlighting the challenge of applying data-hungry architectures to datasets with class imbalance and subtle lesion patterns.

\begin{table*}[!t]
\centering
\caption{Comparative performance in SaNMoD data, for binary classification of DR.}
\label{tab:DR_NoDR}
\resizebox{0.7\textwidth}{!}{%
\begin{tabular}{lccccc}
\hline
\multirow{2}{*}{\textbf{Model}} & \multirow{2}{*}{\textbf{Acc. ($\uparrow$)}} & \multirow{2}{*}{\textbf{B. Acc. ($\uparrow$)}} & \multicolumn{2}{c}{\textbf{F1-score ($\uparrow$)}} & \multirow{2}{*}{\textbf{AUC-PR ($\uparrow$)}} \\ \cline{4-5}
                                &                                             &                                               & \textbf{No DR}          & \textbf{DR}             &                                              \\ \hline
\textbf{VGG16}                  & 0.9138                                      & 0.9117                                        & 0.9229                  & 0.9023                  & 0.8626                                       \\ \hline
\textbf{Densenet121}            & 0.9150                                      & 0.9104                                        & 0.9255                  & 0.9011                  & 0.8722                                       \\ \hline
\textbf{InceptionNetV3}         & 0.9138                                      & 0.9099                                        & 0.9240                  & 0.9004                  & 0.8682                                       \\ \hline
\textbf{EfficientNetB2}         & \textbf{0.9185}                             & \textbf{0.9128}                               & \textbf{0.9292}         & \textbf{0.9040}         & \textbf{0.8820}                              \\ \hline
\textbf{ResNet50}               & 0.9091                                      & 0.9023                                        & 0.9217                  & 0.8917                  & 0.8703                                       \\ \hline
\textbf{ViT}                    & 0.8158                                      & 0.8108                                        & 0.8375                  & 0.7875                  & 0.7257                                       \\ \hline
\end{tabular}%
}
\end{table*}

\begin{table*}[!t]
\centering
\caption{Comparative performance in SaNMoD data, for binary classification of RDR.}
\label{tab:RDR_NonRDR}
\resizebox{0.7\textwidth}{!}{%
\begin{tabular}{lccccc}
\hline
\multirow{2}{*}{\textbf{Model}} & \multirow{2}{*}{\textbf{Acc. ($\uparrow$)}} & \multirow{2}{*}{\textbf{B. Acc. ($\uparrow$)}} & \multicolumn{2}{c}{\textbf{F1-score ($\uparrow$)}} & \multirow{2}{*}{\textbf{AUC-PR ($\uparrow$)}} \\ \cline{4-5}
                                &                                             &                                                & \textbf{Non RDR}         & \textbf{RDR}            &                                               \\ \hline
\textbf{VGG16}                  & 0.7513                                      & 0.7550                                         & 0.6412                   & 0.8097                  & 0.8403                                        \\ \hline
\textbf{Densenet121}            & 0.7487                                      & 0.7424                                         & 0.6275                   & 0.8104                  & 0.8319                                        \\ \hline
\textbf{InceptionNetV3}         & 0.7566                                      & \textbf{0.7614}                                & \textbf{0.6489}          & 0.8138                  & \textbf{0.8443}                               \\ \hline
\textbf{EfficientNetB2}         & \textbf{0.7646}                             & 0.7536                                         & 0.6426                   & \textbf{0.8245}         & 0.8380                                        \\ \hline
\textbf{ResNet50}               & 0.7328                                      & 0.7446                                         & 0.6273                   & 0.7918                  & 0.8349                                        \\ \hline
\textbf{ViT}                    & 0.7275                                      & 0.6712                                         & 0.5339                   & 0.8075                  & 0.7896                                        \\ \hline
\end{tabular}%
}
\end{table*}

The RDR task emphasized clinically actionable cases, having more prominent features such as widespread hemorrhages and venous changes. Table~\ref{tab:RDR_NonRDR} shows the comparative performance of the models on the RDR binary classification task. Here, EfficientNetB2 and InceptionNetV3 achieved the best balance of sensitivity and specificity. Although all CNN models performed comparably, ViT once again underperformed with unstable recall values and poor boundary discrimination. These results confirmed the utility of SaNMoD for benchmarking referral-level classification, while underscoring the difficulty of modeling nuanced lesion boundaries involving limited and imbalanced data.

\begin{table*}[!t]
\centering
\caption{Comparative performance in SaNMoD data, for binary classification of DME.}
\label{tab:DME_NoDME}
\resizebox{0.7\textwidth}{!}{%
\begin{tabular}{lccccc}
\hline
\multirow{2}{*}{\textbf{Model}} & \multirow{2}{*}{\textbf{Acc. ($\uparrow$)}} & \multirow{2}{*}{\textbf{B. Acc. ($\uparrow$)}} & \multicolumn{2}{c}{\textbf{F1-score ($\uparrow$)}} & \multirow{2}{*}{\textbf{AUC-PR ($\uparrow$)}} \\ \cline{4-5}
                                &                                             &                                                & \textbf{No DME}          & \textbf{DME}            &                                               \\ \hline
\textbf{VGG16}                  & \textbf{0.8016}                             & \textbf{0.7805}                                & \textbf{0.8421}          & \textbf{0.7331}         & \textbf{0.6878}                               \\ \hline
\textbf{Densenet121}            & 0.7751                                      & 0.7597                                         & 0.8156                   & 0.7119                  & 0.6464                                        \\ \hline
\textbf{InceptionNetV3}         & 0.7884                                      & 0.7729                                         & 0.8268                   & 0.7279                  & 0.6646                                        \\ \hline
\textbf{EfficientNetB2}         & 0.7804                                      & 0.7652                                         & 0.8200                   & 0.7186                  & 0.6535                                        \\ \hline
\textbf{ResNet50}               & 0.7804                                      & 0.7615                                         & 0.8230                   & 0.7108                  & 0.6553                                        \\ \hline
\textbf{ViT}                    & 0.7381                                      & 0.7188                                         & 0.7880                   & 0.6574                  & 0.5995                                        \\ \hline
\end{tabular}%
}
\end{table*}

Classification of DME posed a distinct challenge due to the localized pathology of macular involvement. Table~\ref{tab:DME_NoDME} demonstrates the performance generated for the binary classification of DME. VGG16 performed best in this task, demonstrating the strength of architectures with strong local feature sensitivity. Other CNNs, including InceptionNetV3 and EfficientNetB2, delivered comparable but slightly lower performance. ViT showed the weakest results, with an AUC-PR less than 0.60, confirming its vulnerability to spatially localized pathologies related to the disease.

\begin{table*}[!t]
\centering
\caption{Comparative performance in SaNMoD data, for multi-class grading of the severity of DR}
\label{tab:DRSeverity}
\resizebox{0.95\textwidth}{!}{%
\begin{tabular}{lcccccccc}
\hline
\multirow{2}{*}{\textbf{Models}} & \multirow{2}{*}{\textbf{Acc. $(\uparrow)$}} & \multirow{2}{*}{\textbf{B. Acc. $(\uparrow)$}} & \multicolumn{5}{c}{\textbf{F1-score $(\uparrow)$}}                                        & \multicolumn{1}{l}{\multirow{2}{*}{\textbf{AUC-PR $(\uparrow)$}}} \\ \cline{4-8}
                                 &                                             &                                                & \textbf{No-DR}  & \textbf{Mild}   & \textbf{Moderate} & \textbf{Severe} & \textbf{PDR}    & \multicolumn{1}{l}{}                                              \\ \hline
\textbf{VGG16}                   & 0.7769                                      & 0.5983                                         & \textbf{0.9253} & 0.4493          & 0.6772            & 0.6122          & 0.3377          & 0.4459                                                            \\ \hline
\textbf{DenseNet121}             & 0.6128                                      & 0.4005                                         & 0.7939          & 0.2105          & 0.5442            & 0.2712          & 0.2041          & 0.3458                                                            \\ \hline
\textbf{InceptionV3}             & 0.7863                                      & 0.6506                                         & 0.9151          & 0.4234          & 0.7206            & 0.4842          & \textbf{0.6569} & 0.4856                                                            \\ \hline
\textbf{EfficientNet B2}         & \textbf{0.7898}                             & 0.5895                                         & 0.9160          & 0.3913          & \textbf{0.7550}   & \textbf{0.6525} & 0.2857          & 0.4569                                                            \\ \hline
\textbf{ResNet50}                & \textbf{0.7898}                             & \textbf{0.6759}                                & 0.9226          & \textbf{0.5207} & 0.7021            & 0.4651          & 0.6497          & \textbf{0.4977}                                                   \\ \hline
\textbf{ViT}                     & 0.5773                                      & 0.4095                                         & 0.8200          & 0.2000          & 0.4300            & 0.2300          & 0.2800          & 0.3786                                                            \\ \hline
\end{tabular}%
}
\end{table*}

Multi-class DR severity grading, summarized in Table \ref{tab:DRSeverity}, was the most challenging task. This was primarily due to the overlap between lesion patterns across adjacent grades alomng with the severe class imbalance. ResNet50 produced the most balanced performance across grades. This was followed by InceptionNetV3 and EfficientNetB2, which leveraged hierarchical and multi-scale representations. DenseNet121 and ViT consistently underperformed, as they struggled to differentiate between subtle lesion progression.

Across all the tasks, in general, the CNN-based architectures consistently outperformed ViT. This reaffirms the importance of spatial inductive biases in handling subtle and localized retinal lesions. InceptionNetV3 offered the best overall balance between performance and efficiency, making it particularly promising for deployment in clinical settings. Severity grading highlighted the typical challenges of DR datasets -- such as class imbalance, morphological overlap, and minority under-representation, that strongly influence model robustness. These findings illustrate that SaNMoD provides both a rich testbed for benchmarking and a realistic reflection of clinical data distributions. The case study confirms its value in developing reliable automated screening pipelines, while emphasizing the need for larger, more  balanced datasets, to unlock the potential of transformer approaches.

\begin{figure}[htbp]
\centering
\includegraphics[width=0.7\textwidth]{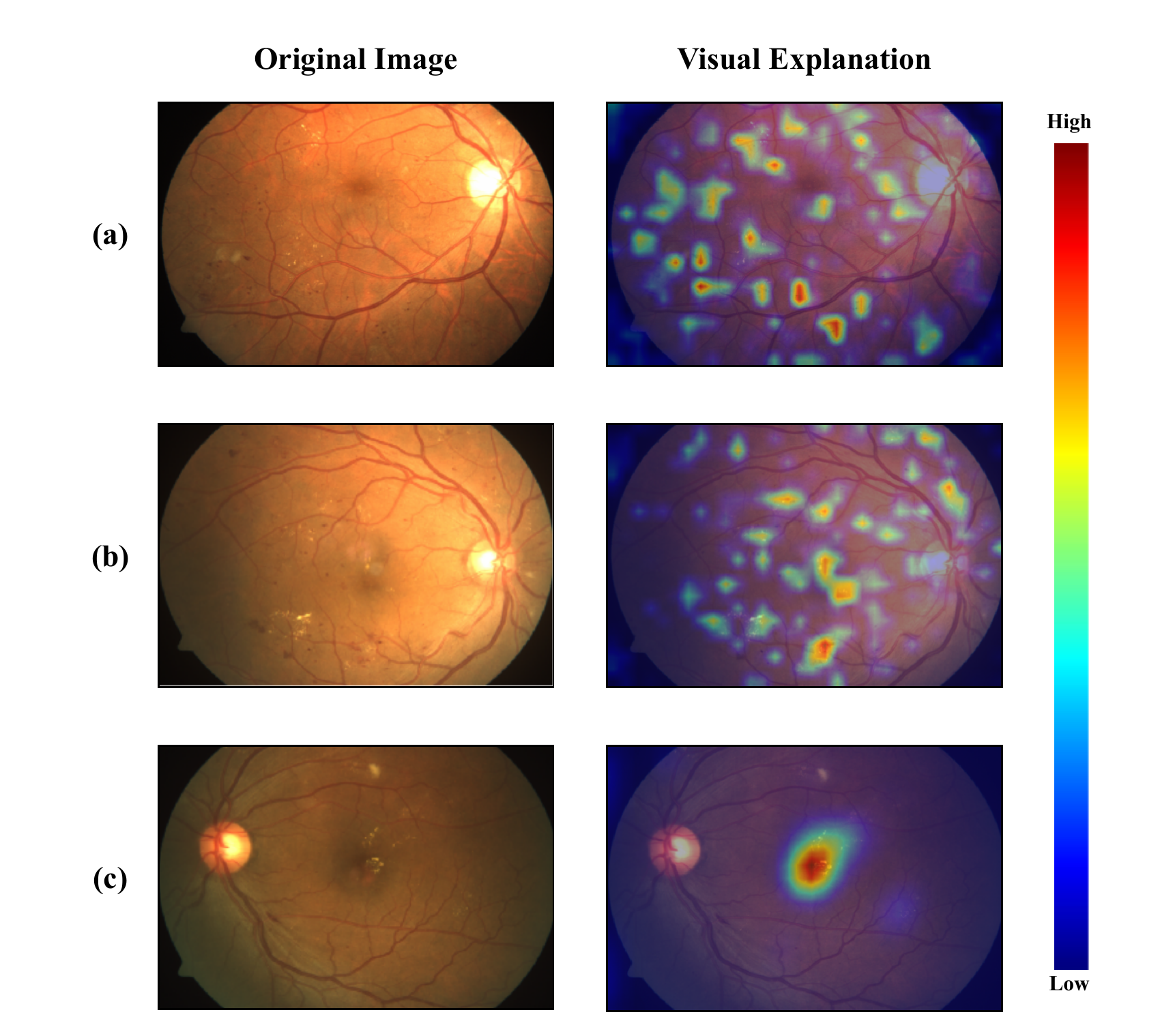}
\caption{Representative Grad-CAM visualization, highlighting lesion-specific activation for (a) DR, (b) RDR, and (c) DME binary classification tasks.}
\label{visualization}
\end{figure}

Grad-CAM visualization \cite{selvaraju2017grad} was used to interpret model decisions across the three binary classification tasks. Fig.~\ref{visualization} shows representative examples, with the original fundus images in the first column and the corresponding activation maps in the second. In DR detection [Row (a)], high-activation regions aligned with microaneurysms and bright reflective spots, consistent with clinical markers. A patch of blot hemorrhage in the lower left quadrant also exhibited strong activation. For the RDR classification [Row (b)], the models emphasized densely distributed microaneurysms and exudates near the macula and superior arcade. For DME detection [Row (c)], activation maps concentrated on bilateral macular exudates; particularly, in the lower central hemisphere and closely matching with clinical signs. These visualizations confirmed that models trained on SaNMoD emphasized clinically relevant lesion clusters, mostly near the fovea, while downplaying peripheral artifacts.

\subsection{Insights}
\label{subsec:keytakeaways}

The key takeaway from this case study emphasizes how the properties of a dataset influence model outcomes and provide guidance in aligning with suitable architectures. These mainly encompass the following.

\begin{itemize}
    \item Class imbalance, subtle lesion morphology, and overlapping between severity grades constrain the performance of a model. This reflects the challenges in most DR datasets.  
    \item Lesion-centric datasets favor CNNs with spatial inductive biases, while scarce and imbalanced data hinder the performance of transformer-based models.  
    \item Diversity in lesion sizes in the datasets demands multi-scale feature extraction. This is a characteristic of InceptionNetV3, for stabilized performance in binary classification tasks.  
    \item Although RDR labels create a clinically actionable benchmark, the limited samples in datasets reduce their reliability in precisely modeling intricate lesion boundaries.  
    \item Localized pathologies, such as in DME, require high-resolution annotations to enable models sensitive to fine-grained features to perform effectively. VGG, with its strong inductive bias, proved effective in this.
    \item Lesion annotations strengthen interpretability, as observed in Grad-CAM visualizations aligning activations with clinically relevant retinal regions.  
    \item Balanced and richly annotated datasets enable robust benchmarking to support meaningful clinical translation. 
\end{itemize}

\section{Conclusion}
\label{sec:conclusion}

This article presented a comprehensive review of fundus image datasets for the screening of DR, focusing on their usability for automated systems developed using deep learning. The analysis showed that repositories such as Messidor, EyePACS, DDR, IDRiD, and SaNMoD enabled progress in binary classification, severity grading, lesion localization, and screening for multiple diseases. However, they remained constrained by class imbalance, limited demographic coverage, and inconsistent annotation protocols. DL models are critically dependent on the quality of annotations and labels in the dataset, to learn robust and clinically reliable features. The shortcomings of the datasets, in these areas, reduced their generalizability across populations; thereby, and hindering clinical translation.

Analysis in DR datasets established that characteristics such as lesion subtlety, annotation richness, and severity distribution directly influence model robustness, generalization, and interpretability. These patterns, evident in SaNMoD, were found in other public repositories. It underscores the central role of the quality of datasets in shaping the outcome of a model.

Efforts are needed to prioritize standardized annotation frameworks, inclusion of lesion annotation, and longitudinal data, along with the integration of multiple disease labels with relevant clinical metadata. Lightweight models, trained on such datasets, can improve diagnostic efficiency; particularly, in resource-constrained environments. Finally, well-curated and representative datasets serve as the basis for explainable and clinically reliable systems that can aim to reduce the global burden of blindness related to DR.


\section*{Use of AI Tools}
The authors declare that no Artificial Intelligence (AI) tools, including Large-Language Models (LLMs), were used for content generation in this manuscript.

\section*{Funding Information}
The research  was supported by the J. C. Bose National Fellowship of S. Mitra, under
Grant Number JCB/2020/000033.

\section*{Conflict of Interests}
The authors declare no potential conflict of interests.

\bibliographystyle{unsrt}  
\bibliography{references}

\end{document}